\documentclass[showkeys,eqnum,showpacs,prd]{revtex4}
\usepackage[dvips]{graphicx}
\usepackage{color}
\usepackage{amsmath}
\begin{document}

\title{Notes Concerning "On the Origin of Gravity and the Laws of Newton" by E. Verlinde}

\author{Jarmo M\"akel\"a} 

\email[Electronic address: ]{jarmo.makela@puv.fi}  
\affiliation{Vaasa University of Applied Sciences, Wolffintie 30, 65200 Vaasa, Finland}

\begin{abstract} 
 
  We point out that certain equations which, in a very recent paper written by E. Verlinde, are
postulated as a starting point for a thermodynamical derivation of classical gravity, are
actually consequences of a specific microscopic model of spacetime, which has been published earlier.

\end{abstract}


\maketitle

\section{Introduction}

  A remarkable paper was published very recently by Verlinde. \cite{yksi} In that paper it was argued, 
among other things, that Newton's second law of dynamics and Newton's universal law of 
gravitation arise naturally and unavoidably in a theory in which space is emergent through 
a holographic scenario. \cite{kaksi} In these notes we shall focus our attention to Verlinde's derivation of 
Newton's law of gravitation. 

  The starting point of Verlinde's derivation was an assumption that whenever we have a sphere with 
area $A$, the sphere acts as a storage device for information such that the total number of bits of 
information stored on the surface of the sphere is
\begin{equation}
N = \frac{Ac^3}{G\hbar}.
\end{equation}
Verlinde then identified as the energy $E$ of a system the rest energy $Mc^2$ of the mass inside the sphere,
and he assumed that the energy is divided evenly over the bits $N$ such that
\begin{equation}
E = \frac{1}{2}Nk_BT.
\end{equation}
If we put $E = Mc^2$ in Eq. (2), identify $T$ as the Unruh temperature
\begin{equation}
T_U := \frac{1}{2\pi}\frac{\hbar a}{k_Bc}
\end{equation}
of an observer with a proper acceleration $a$, and use the fact that the area of a two-sphere with radius $r$ is
\begin{equation}
A = 4\pi r^2,
\end{equation}
we find, by means of Newton's second law $F = ma$, that the force exerted by a point-like mass $M$ on a mass $m$
at a distance $r$ is:
\begin{equation}
F = G\frac{mM}{r^2},
\end{equation}
which is Newton's universal law of gravitation. A very similar derivation 
of Eq. (5) was also discovered by Padmanabhan. \cite{kolme}

  The validity of Verlinde's derivation hinges on Eqs. (1) and (2), (Eqs. (3.10) and (3.11) in Verlinde's paper)
together with an identification of the absolute temperature $T$ as the Unruh temperature $T_U$. Are 
Eqs. (1) and (2) valid? In which sense can we say that the temperature $T$ may be identified as the 
Unruh temperature $T_U$? 

   The questions raised above could be addressed in a precise fashion, if we had a concrete microscopic model of 
spacetime in our disposal. Using that model we should be able to calculate the partition function of spacetime, 
which, in turn, would yield an expression to the entropy - and therefore information - stored on its two-dimensional
surfaces. We should also be able to calculate the dependence of the average energy associated with those surfaces 
on the given temperature $T$. Finally, we should be able to gain better understanding on the role played by the
Unruh temperature $T_U$ in the thermodynamics of spacetime. Even if our model were far from correct, it could 
possess features, which are generic to every plausible microscopic model of spacetime, thus providing a useful 
starting point for more advanced models.

   Such a model indeed exists, and it was developed in details in Refs. \cite{nelja, viisi, kuusi}. In what 
follows, we shall outline that model briefly, and comment on its bearing on Verlinde's approach.
  
   In Refs. \cite{nelja, viisi, kuusi} spacetime was modelled by a specific graph with particular objects on 
its vertices, which were called, for certain
reasons, as "microscopic quantum black holes". The idea was to reduce all properties of spacetime to the quantum 
states of those holes. At macroscopic scales the quantum black holes are assumed to generate areas for spacelike 
two-surfaces of spacetime. More precisely, the possible eigenvalues of the areas of spacelike two-surfaces are 
assumed to be of the form:
\begin{equation}
A = \alpha(n_1 + n_2 + ... + n_N)\ell_{Pl}^2,
\end{equation}
where $\alpha$ is a numerical coefficient to be determined later, $N$ is the number of quantum black holes
on the surface, and
\begin{equation}
\ell_{Pl} := \sqrt{\frac{\hbar G}{c^3}} \approx 1.6\times 10^{-35}m
\end{equation}
is the Planck length. The numbers $n_1, n_2, ..., n_N$ are non-negative integers, which label the quantum states 
of the $N$ quantum black holes lying on the surface. In other words, each quantum black hole on the surface is 
assumed to contribute to the surface an area, which is of the form:
\begin{equation}
A_n = \alpha n\ell_{Pl}^2,
\end{equation}
where $n = 0, 1, 2,...$. The microstates of the surface are assumed to be uniquely determined by the combinations
of the non-vacuum $(n\ne 0)$ area eigenstates of the holes, whereas the microstates corresponding to the same total 
area of the surface belong to the same macroscopic state of the surface.

   A specific attention in the model was paid to the so called {\it acceleration surfaces}. In very broad terms, 
acceleration surface may be described as a spacelike two-surface propagating in spacetime such that each point 
of the surface accelerates with the same constant proper acceleration $a$ in  a direction orthogonal to the 
surface (For a mathematically precise definition, see Refs. \cite{viisi} and \cite{kuusi}). An example of an acceleration surface is
 provided by a $t = constant$ slice of a timelike hypersurface
$r = constant$ of the Schwarzschild spacetime equipped with the Schwarzschild coordinates $r$ and $t$. Indeed, 
each point of such a slice accelerates with the same proper acceleration
\begin{equation}
a = (1 - \frac{2M}{r})^{-1/2}\frac{M}{r^2}.
\end{equation}
In the Newtonian limit and SI units this gives:
\begin{equation}
a = G\frac{M}{r^2},
\end{equation}
which is the acceleration of particles in a free fall in the gravitational field created by a point-like mass $M$.

  One of the reasons for defining the concept of acceleration surface is that with acceleration surfaces it is 
possible to associate the concept of energy in a natural manner. In the model of Refs. \cite{nelja, viisi, kuusi}
the macroscopic energy 
of an  acceleration surface with proper acceleration $a$ and area $A$ was defined to be, from the point of view 
of an observer at rest with respect to the surface,
\begin{equation}
E := \frac{c^2aA}{4\pi G}.
\end{equation}
If one takes the acceleration surface to be  a two-sphere with radius $r$ surrounding a point-like mass $M$, and
identifies $E$ as $Mc^2$, one immediately finds, in the Newtonian limit, that the gravitational force exerted by
the mass $M$ on a mass $m$ is
\begin{equation}
F = G\frac{Mm}{r^2}.
\end{equation}
In other words, Newton's universal law of gravitation is, in our model, a consequence of our
macroscopic definition of energy, rather than an outcome of the microscopic properties of the constituents
of spacetime.

   Using Eqs. (6) and (11) we find that the energies of an acceleration surface are of the form:
\begin{equation}
E_n = n\frac{\alpha\hbar a}{4\pi c},
\end{equation}
where
\begin{equation}
n := n_1 + n_2 + ... + n_N.
\end{equation}
Hence we are able to calculate the {\it partition function}
\begin{equation}
Z(\beta) := \sum_{n=1}^\infty g(E_n)e^{-\beta E_n}
\end{equation}
of the acceleration surface. It follows from our definitions of the microscopic and the macroscopic states 
of spacelike two-surfaces of spacetime that the degeneracy $g(E_n)$ of a state of an acceleration surface 
with energy $E_n$ is simply the number of the possible ways of writing the positive integer $n$ as a sum of at most $N$
positive integers. More precisely, $g(E_n)$ is the number of the ordered strings $(n_1, n_2, ..., n_m)$, where
$n_1, n_2, ..., n_m \in \lbrace 1, 2, 3, ...\rbrace$ and $1\le m\le N$ such that
\begin{equation}
n_1 + n_2 + ... + n_m = n.
\end{equation}

  A rather involved calculation yields an explicit -and surprisingly simple- analytical expression for the 
partition function $Z(\beta)$. \cite{kuusi} Using the partition function $Z(\beta)$  one may obtain explicit expressions
both for the average energy
\begin{equation}
E(\beta) = -\frac{\partial}{\partial\beta}\ln Z(\beta)
\end{equation}
and the entropy
\begin{equation}
S(\beta) = k_B\biggl[\ln Z(\beta) - \beta\frac{\partial}{\partial\beta}\ln Z(\beta)\biggr]
\end{equation}
of an acceleration surface in a given temperature $T$. The key role in the results obtainable from these 
expressions is played by the {\it characteristic temperature}
\begin{equation}
T_C := \frac{\alpha}{4\pi\ln 2}\frac{\hbar a}{k_B c}
\end{equation}
of the acceleration surface. It also turns out useful to define a quantity
\begin{equation}
\bar{E}(\beta) := \frac{E(\beta)}{N},
\end{equation}
which tells the average energy per a constituent of the acceleration surface.

   The results of our model relevant to our comments on Verlinde's paper are the following: \cite{kuusi}
\begin{subequations}
\begin{eqnarray}  
    \lim_{N\rightarrow\infty}\bar{E}(T) &=& 0,\,\,\,\,\,\forall\, T < T_C,\\
      \bar{E}(T_C) &=& k_BT_C\ln 2,\\
\frac{d\bar{E}(T)}{dT}\vert_{T=T_C} &=& \frac{1}{6}k_B(\ln 2)^2N + O(1),\\
S(E) &=& \frac{2\pi k_Bc}{\hbar a}E\,\,\,\,\,\,\,\forall\, E \le 2Nk_BT_C.
\end{eqnarray}
\end{subequations}

  Eq. (21a) (Eq. (4.7) in Ref. \cite{kuusi}) implies that in the large $N$ limit the microscopic quantum black 
holes on the acceleration surface are, in average, in vacuum, when $T < T_C$,
whereas Eq. (21c) (Eq. (4.23) in Ref. \cite{kuusi}) implies that when $T \approx T_C$, the curve 
$\bar{E} = \bar{E}(T)$ is practically vertical (See Fig. 1). The verticality
of the curve $\bar{E} = \bar{E}(T)$, when $T \approx T_C$ indicates a {\it phase
transition} at the temperature $T = T_C$. A detailed analysis shows that during
this phase transition the quantum black holes on the acceleration surface jump,
in average, from the vacuum, where $n = 0$, to the second excited states, where
$n = 2$. When $T >> T_C$, one finds that $E \approx Nk_BT$ as a very good 
approximation. \cite{kuusi}

\begin{figure}[htb!]
\begin{center} 
\includegraphics{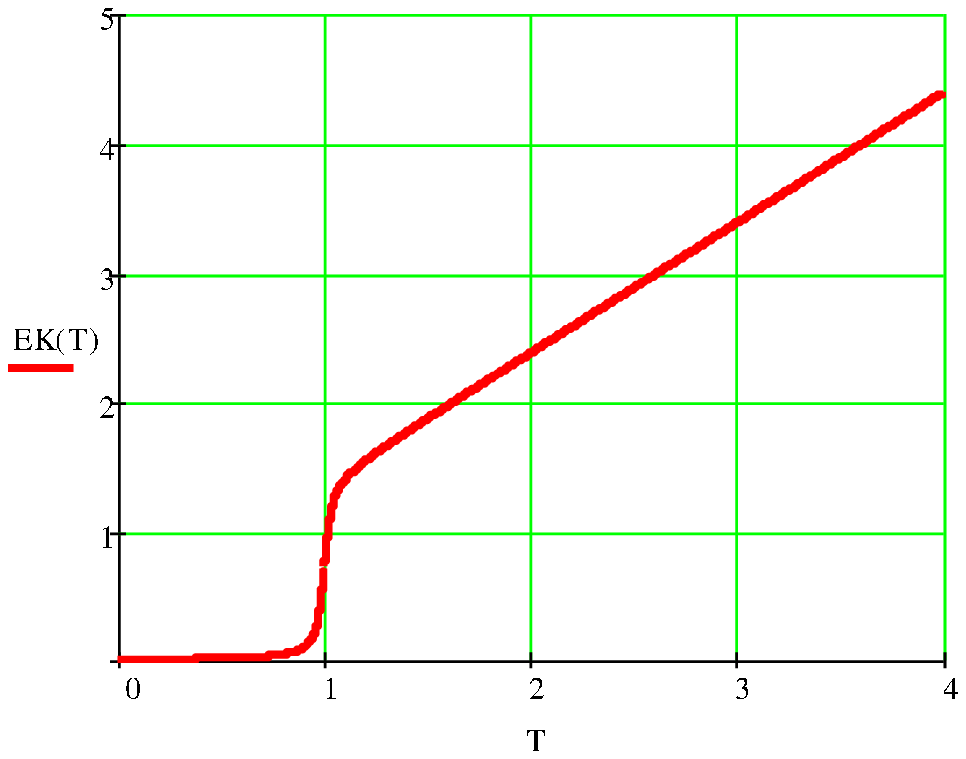}
\caption{The average energy ${\bar E}$ $(= EK(T))$ of an acceleration surface per a hole as a function of the absolute 
temperature $T$, when the number of the holes on the surface is $N = 100$. The absolute temperature $T$ has been 
expressed in the units of $T_C$, and the average energy ${\bar E}$ in the units of $k_BT_C$. We have chosen $\alpha = 2\ln2$.
 If $T < T_C$, 
${\bar E}$ is effectively zero, which means that the black holes on the acceleration surface are in vacuum. When
$T = T_C$, the curve ${\bar E} = {\bar E}(T)$ is practically vertical, which indicates a phase transition at the 
temperature $T = T_C$. During this phase transition the black holes on the acceleration surface are excited from
the vacuum to the second excited states. The latent heat per a hole corresponding to this phase transition is
${\bar L} = 2(\ln 2)k_BT_C \approx 1.4k_BT_C$. When $T > T_C$, the curve ${\bar E} = {\bar E}(T)$ is approximately 
linear.
  \label{fig:Figure1}} 
\end{center}
\end{figure}

    The fact that the black holes constituting the acceleration surface are in 
vacuum, whenever $T < T_C$ implies that the characteristic temperature $T_C$ is
the lowest possible temperature which an observer at rest with respect to the 
acceleration surface may measure, and therefore $T_C$ may be identified as the
Unruh temperature $T_U$ of Eq. (3) measured by an accelerated observer. Indeed, 
the characteristic temperature $T_C$ coincides with the Unruh temperature $T_U$,
if we fix the undetermined constant $\alpha$ in Eq. (6) as:
\begin{equation}
\alpha = 2\ln 2.
\end{equation}
So we see that our model predicts the Unruh effect: When the matter fields are 
in thermal equilibrium with spacetime, an accelerated observer will detect thermal
radiation with a characteristic temperature, which is proportional to the proper
acceleration $a$ of the observer.

   Eq. (21d) (Eq. (5.25) in Ref. \cite{kuusi})holds with the choice of Eq. (22) for the constant $\alpha$. Using
Eqs. (21b) (Eq. (4.22) in Ref. \cite{kuusi})and (21d) one finds that at the Unruh temperature $T_U = T_C$ the entropy $S$ of
an acceleration surface may be written as:
\begin{equation}
S(T_U) = Nk_B\ln 2.
\end{equation}
Because the relationship between $H$, the number of bits of information stored
on a system, and its entropy $S$ is, in general:
\begin{equation}
H = \frac{S}{k_B\ln2},
\end{equation}
we find that the amount of information carried by the acceleration surface at the
Unruh temperature $T_U$ is:
\begin{equation}
H = N.
\end{equation}
In other words, each quantum black hole on the acceleration surface carries, in average,
exactly one bit of information. This is a very nice result, and using Eq. (11) we observe 
that at the Unruh temperature $T_U$ the relationship between the area $A$ and the integer
$N$ is:
\begin{equation}
N = \frac{1}{2\ln 2}\frac{Ac^3}{G\hbar}.
\end{equation}
Moreover, Eq. (21b) implies that at the Unruh temperature $T_U$ the energy of the 
acceleration surface is:
\begin{equation}
E = Nk_BT_U\ln 2.
\end{equation}
As one may observe, up to a numerical coefficient $2\ln 2$, Eqs. (26) and (27), respectively,
are identical to Eqs. (1) and (2). 

  In these notes we have shown that Eqs. (1) and (2) which, in Verlinde's paper, \cite{yksi} constitute
a starting point for his derivation of Newton's universal law of gravitation from the 
thermodynamical properties of spacetime follow, up to an unimportant numerical factor, 
from a specific microscopic model of spacetime. \cite{nelja, viisi, kuusi} In that model 
spacetime is assumed to be a graph
with objects called as "microscopic quantum black holes" on its vertices. The attention is 
focused at the so called acceleration surfaces, and each quantum black hole on an acceleration
surface is assumed to contribute to the surface an area which, up to a certain numerical 
coefficient, is an integer times the square of the Planck length. According to our model 
the Unruh temperature $T_U$ of an accelerated observer is the lowest possible temperature
which an accelerated observer may measure, provided that the matter fields are, from the 
point of view of the observer, in thermal equilibrium with spacetime. At the Unruh 
temperature the acceleration surface performs a phase transition, where the quantum black
holes on the surface jump, in average, from the vacuum to the second excited states. Eqs.
(1) and (2) hold in the Unruh temperature, as it was conjectured by Verlinde.

  The results obtainable from our model provide support for Verlinde's results. Indeed, it
is possible to construct a concrete microscopic model of spacetime, where equations acting
as a starting point of Verlinde's approach follow as special cases. However, it should be 
noted that the logic in Verlinde's paper and in our model is somewhat different: In 
Verlinde's paper one assumes that the information  and the energy carried by the constituents
of a two-sphere surrounding a point-like mass $M$ obey, in the Unruh temperature, Eqs. (1) and 
(2). An identification of $Mc^2$ as the total energy associated with that two-sphere implied,
through Eqs. (1) and (2), Newton's universal law of gravitation. In our model, in turn, one 
defines energy as an entirely macroscopic concept. In that definition a certain relationship
between the total energy, proper acceleration and the area of an acceleration surface is 
postulated and the postulate in question implies, in the Newtonian limit, Newton's law of 
gravitation. In the microscopic level the postulate implies, at the Unruh temperature and up
to a certain numerical coefficient, Eqs. (1) and (2) for the constitutes of the two-sphere.

  Taken as a whole, our model and Verlinde's approach may be seen as complementary: In 
Verlinde's approach Newton's  universal law of gravitation is a consequence of certain thermal
and entropic properties of the constituents of spacetime, whereas in our model those properties
are, in a sense described above, necessary consequences of Newton's law of gravitation. If the
idea of gravity as an emergent effect, rather than as a fundamental force turns out to be correct,
we may currently be in a somewhat similar position as were the founders of quantum mechanics
and atomic physics about 100 years ago. Instead of attempting to understand the microstructure 
of matter, however, we should, in this time, attempt to understand the microstructure of spacetime 
itself. It remains to be seen, whether any features of our still very preliminary microscopic model
of spacetime will survive in more advanced models. Be that as it may, exciting times will lie
ahead.

   \end{document}